\documentclass[proceedings]{JHEP3}
\usepackage{amssymb}
\usepackage{amsfonts}
\usepackage{amsmath}
\usepackage{epsfig,multicol}

\newbox\mybox

\newcommand\fverb{\setbox\mybox=\hbox\bgroup\verb}
\newcommand\fverbdo{\egroup\medskip\noindent\fbox{\unhbox\mybox}\ }
\newcommand\fverbit{\egroup\item[\fbox{\unhbox\mybox}]}
\conference{A Q-operator for the twisted XXX model}
\abstract{Taking the isotropic limit $\Delta\rightarrow 1$ in a recent 
representation theoretic construction of Baxter's Q-operators for the XXZ model with
quasi-periodic boundary conditions we obtain new results for the XXX model. We show 
that quasi-periodic boundary conditions are needed to ensure convergence of the 
Q-operator construction and derive a quantum Wronskian relation which implies two 
different sets of Bethe ansatz equations, one above the other below the "equator" of total
spin $S^z=0$.
We discuss the limit to periodic boundary conditions at the end and explain how this 
construction might be useful in the context of correlation functions on the infinite 
lattice. We also identify a special subclass of solutions to the quantum Wronskian for 
chains up to a length of 10 sites and possibly higher.}
\input{tcilatex}

\title{A Q-operator for the twisted XXX model}
\author{Christian Korff \\
Centre for Mathematical Science, City University, \\
Northampton Square, London EC1V 0HB, UK\\
E-mail: \email{c.korff@city.ac.uk}
}

\begin{document}

\section{Introduction}

Historically Baxter's Q-operator was introduced as substitute method for the
coordinate Bethe ansatz in solving the eight-vertex model \cite%
{Bx72,Bx73,Bx82}, but has more recently seen wider applications in the field
of integrable systems e.g. \cite{PG,KLWZ97,BLZ97,KS98,Sk00,FKV01} making it an
important and almost universal tool. To motivate his technique Baxter first
discussed the concept of the Q-operator in the context of the six-vertex or
XXZ model, where a direct comparison with the coordinate Bethe ansatz can be
made. While our primary interest in this article will be the XXX model it is
helpful to consider first the anisotropic or XXZ case. Denote by $t$ the
transfer matrix then the $Q$-operator is implicitly defined through the
functional equation%
\begin{equation}
t(u)Q(u)=\chi (u-\tfrac{1}{2})^{M}Q(u+1)+\chi (u+\tfrac{1}{2})^{M}Q(u-1)
\label{TQintro}
\end{equation}%
where $\chi $ is an explicitly known function (the quantum determinant) and $%
M$ the number of lattice columns, respectively the number of sites in the
spin-chain. In addition, to this relation, known as $TQ$ equation, one
usually requires a number of properties such as \textquotedblleft
analyticity\textquotedblright\ of the $Q$-operator in the spectral variable $%
u$ and that $[T(u),Q(u^{\prime })]=[Q(u),Q(u^{\prime })]=0$ for an arbitrary
pair $u,u^{\prime }\in \mathbb{C}$. The latter commutation relations allow
one to discuss the TQ equation on the level of eigenvalues and this is where
one makes contact with the coordinate Bethe ansatz \cite{Bethe} which
determines the spectrum of the transfer matrix in terms of the solutions $%
\{v_{k}\}_{k=1}^{n}$ to the Bethe ansatz equations \cite{Li67,Su67},%
\begin{equation}
\left( \frac{\sinh \gamma (v_{j}-i/2)}{\sinh \gamma (v_{j}+i/2)}\right)
^{M}=\prod_{k\neq j}\frac{\sinh \gamma (v_{j}-v_{k}-i)}{\sinh \gamma
(v_{j}-v_{k}+i)},\qquad j=1,2,...,n=M/2-S^{z}\ .  \label{xxzBAE}
\end{equation}%
Here $\gamma $ is the crossing or coupling parameter of the six-vertex model
and $S^{z}\geq 0$ the total spin-operator. Postulating that the eigenvalues
of the $Q$-operator are of the form \cite{Bx82}%
\begin{equation}
Q(u)=\tprod_{j=1}^{n}\frac{\sinh \gamma (u-v_{j})}{\sinh \gamma }
\end{equation}%
the outcome of the coordinate Bethe ansatz then implies the $TQ$ relation (%
\ref{TQintro}), which is the starting point for the construction of the
operator $Q$. Note that this line of argument is based on the essential
assumption that the coordinate Bethe ansatz yields a complete set of
eigenstates of the transfer matrix with a finite set of Bethe roots $v_{j}$.
It is this assumption, which has to be treated with care in the isotropic
limit $\gamma \rightarrow 0$ yielding the XXX model\footnote{%
Similar problems occur for the XXZ model at roots of unity \cite{FM01,Bx02}
due to a partial loop algebra symmetry \cite{DFM}.}.

The transfer matrix of the XXX model as well as the associated Heisenberg
spin-chain are $sl_{2}$ symmetric, whence their eigenspaces decompose into $%
sl_{2}$ modules. As is well known the finite solutions to the XXX Bethe
ansatz equations (first derived by Bethe in \cite{Bethe} albeit in a
different form), 
\begin{equation}
\left( \frac{v_{j}-i/2}{v_{j}+i/2}\right) ^{M}=\prod_{k\neq j}\frac{%
v_{j}-v_{k}-i}{v_{j}-v_{k}+i}\ ,  \label{xxxBAE}
\end{equation}%
now only yield the highest weight vectors in each $sl_{2}$ module \cite{FT81}%
. The remaining states within each module are obtained through the action of
the symmetry algebra and have been referred to as \textquotedblleft
non-regular\textquotedblright\ Bethe states as they involve
\textquotedblleft infinite rapidities\textquotedblright\ in the particular
parametrization used in (\ref{xxxBAE}); see e.g. \cite{Deg01} for a
discussion how to recover the non-regular Bethe states through a limiting
procedure. Thus, the obvious ansatz 
\begin{equation}
Q(u)=\tprod_{j=1}^{n}(u-v_{j})  \label{naive}
\end{equation}%
for the eigenvalues of an XXX $Q$-operator becomes problematic due to the
presence of \textquotedblleft infinite rapidities\textquotedblright , or
more precisely not all states correspond to finite solutions of the Bethe
ansatz equations (\ref{xxxBAE}). Clearly, there are ways out of this
dilemma, either by choosing a different parametrization such that all
rapidities stay finite (this is for instance the case in the coordinate
Bethe ansatz, where the non-regular Bethe states correspond to the case that
multiple quasi-momenta vanish), or by (continuously) breaking the $sl_{2}$
symmetry in such a manner that the assumption on the completeness of the
Bethe ansatz becomes applicable again.

In this work we shall do the latter by introducing quasi-periodic boundary
conditions, see e.g. \cite{BY61,dV84,ABB88,YF92}. This has the advantage
that all relevant algebraic properties needed for the quantum inverse
scattering method \cite{FST79} stay intact and that we can take at the very
end the limit to periodic boundary conditions making contact with previous
investigations of Q-operators for the XXX model. Of particular interest will
be aspects which are not accessible through the coordinate Bethe ansatz,
namely the existence of two linearly independent solutions, say $Q^{\pm }$,
to the TQ equation and, closely related with this question, the derivation
of the following quantum Wronskian identity%
\begin{equation}
\frac{\omega Q_{\omega }^{+}(u-\frac{i}{2})Q_{\omega }^{-}(u+\frac{i}{2}%
)-\omega ^{-1}Q_{\omega }^{+}(u+\frac{i}{2})Q_{\omega }^{-}(u-\frac{i}{2})}{%
\omega -\omega ^{-1}}=\chi (u)\ ,  \label{qW}
\end{equation}%
which is a new result. Here $\omega =\exp (i\phi )$ is the twist parameter
associated with the quasi-periodic boundary conditions and $\chi $ is the
aforementioned quantum determinant, but now of the XXX model. Since the
latter is explicitly known, e.g. $\chi (u)=u^{M}$ for the homogeneous case,
one can employ the quantum Wronskian (\ref{qW}) rather than the
generalization of the Bethe ansatz equations (\ref{xxxBAE}) to twisted
boundary conditions when solving the model. Namely, making the ansatz (which
will be justified through our construction of $Q_{\omega }^{\pm }$ in the
text) 
\begin{equation}
Q_{\omega }^{+}(u)=\tprod_{j=1}^{n}(u-v_{j}^{+})\qquad \text{and\qquad }%
Q_{\omega }^{-}(u)=\tprod_{j=1}^{M-n}(u-v_{j}^{-}),\qquad n=\frac{M}{2}-S^{z}
\end{equation}%
for the eigenvalues of the two solutions to the TQ equation, the roots $%
v_{j}^{\pm }=v_{j}^{\pm }(\omega )$ are determined through (\ref{qW}). Here $%
S^{z}$ denotes the total spin component in the direction singled out by the
quasi-periodic boundary conditions. Note that upon setting $u=v_{j}^{\pm
}+i/2,v_{j}^{\pm }-i/2$ the identity (\ref{qW}) implies two different sets
of Bethe ansatz equations, one above, the other one below the equator $%
S^{z}=0$. Due to the quasi-periodic boundary conditions, $\omega \neq 1$,
the Bethe roots $v_{j}^{\pm }$ are all finite and the number of solutions
matches the dimension of each fixed spin-sector signaling completeness;
compare for example with the discussion in \cite{TV95}.

As discussed above this ceases to be true in the limit $\omega \rightarrow 1$
corresponding to periodic boundary conditions. From (\ref{qW}) we infer that
this limit might indeed be singular unless the numerator and denominator
vanish simultaneously. We will compare the outcome of this article with the
findings for periodic boundary conditions by Pronko and Stroganov \cite{PS99}%
, who have presented a similar quantum Wronskian without the denominator at $%
\omega =1$ and a different degree for the second solution $Q^{-}$, namely $%
\deg Q^{-}=M-n+1$. Their Wronskian relation can be numerically solved but
the resulting number of solutions is in general much smaller then the
dimension of the state space $\binom{M}{n}$. In light of the previous
remarks on the $sl_{2}$ symmetry this is not surprising as their solutions
only yield the highest weight vectors in each module. Taking the limit $%
\omega \rightarrow 1$ in the explicit solutions to (\ref{qW}) for small
chains we indeed find that of those solutions $Q_{\omega }^{\pm }$ which
stay finite, \emph{both} approach the $Q^{+}$ solution of Pronko and
Stroganov. We shall comment on this in more detail in the text, see section
5.2.

The appearance of singularities in the limit of periodic boundary conditions
can also be understood from the explicit construction of the Q-operator for
twisted boundary conditions. The latter is given as the trace of a monodromy
matrix with infinite-dimensional auxiliary space. In order to obtain a
well-defined object one must ensure convergence of the trace. As we will see
in the text this actually \emph{requires} the introduction of quasi-periodic
boundary conditions. Previous constructions of $Q$-operators for the XXX
spin-chain \cite{De99,Pr00,De05} have been for periodic boundary conditions
only, where $Q$ has been represented as an integral kernel (see also \cite%
{DKK05} for a related XXZ construction).

In contrast the limit of the transfer matrix from quasi-periodic to periodic
boundary conditions is well-defined. In fact, this applies to all higher
spin transfer matrices which can be expressed in terms of $Q_{\omega }^{\pm
} $ as follows,%
\begin{equation}
t(u;x)=\lim_{\omega \rightarrow 1}\frac{\omega ^{x}Q_{\omega }^{+}(u-\frac{ix%
}{2})Q_{\omega }^{-}(u+\frac{ix}{2})-\omega ^{-x}Q_{\omega }^{+}(u+\frac{ix}{%
2})Q_{\omega }^{-}(u-\frac{ix}{2})}{\omega -\omega ^{-1}}\ .  \label{txintro}
\end{equation}%
When $x=n\in \mathbb{N}_{>0}$ the function $t(u;x=n)$ gives the spectrum of
the transfer matrix with spin $s=(n-1)/2$ in the auxiliary space. However,
if we take $x$ to be an arbitrary complex parameter, the resulting spectrum
belongs to a generalized transfer matrix used in the discussion of
correlation functions for the infinite chain \cite%
{BJMSTxxx,BJMSTxxz,BJMSTxyz,BJMSTxxx2}. This result is the analogue of a
previous discussion for the XXZ model \cite{CKQ3,CKQ7} and the discussion
presented here is in accordance with these earlier results for the more
general case when $\gamma \neq 0$. At the moment there appears to be no
construction of an Q-operator for $\omega =1$ which allows to define (\ref%
{txintro}). This is one of the main reasons for the construction presented
in this paper.

In section 2 the basic definitions of the XXX model and its fusion hierarchy
is stated. Section 3 contains the construction of the Q-operator which is
simply the isotropic limit ($\gamma \rightarrow 0$) of earlier constructions
for the XXZ model \cite{CKQ7}. We briefly address the aforementioned
conditions for convergence due to an infinite-dimensional auxiliary space
and state the relevant functional equations with the transfer matrix. We
omit most proofs for those results which readily follow from taking the
isotropic limit in the XXZ construction. For instance, the eigenvalues of
the Q-operator are discussed by making contact with the algebraic Bethe
ansatz discussion in \cite{CKQ3}. By comparison with the analogous results
for the XXZ model it is shown that the Q-operator factorizes into two
linearly independent solutions to Baxter's TQ-equation. We discuss how they
are related via spin-reversal. The relation with the fusion hierarchy and
its analytic continuation (\ref{txintro}) to \textquotedblleft complex
spin\textquotedblright\ is presented in section 4. Section 5 gives the
quantum Wronskian relation between the two independent solutions to Baxter's
TQ equation, which is then compared against the one of Pronko and Stroganov 
\cite{PS99}. A special subset of solutions to the twisted quantum Wronskian (%
\ref{qW}) is also discussed based on numerical results for chains of even
length $\leq 10$. Their associated Bethe roots obey identities which imply
(and are therefore more fundamental than) the Bethe ansatz equations. The
conclusions are stated in section 6.

\section{Definitions}

Let us start by introducing our conventions for the definition of the XXX
model. Denote by $\{\sigma ^{x}=\sigma ^{1},\sigma ^{y}=\sigma ^{2},\sigma
^{z}=\sigma ^{3}\}$ the Pauli matrices acting on $\mathbb{C}^{2}$ and let $%
\mathbb{P}$ be the permutation operator, $\mathbb{P}(v\otimes w)=w\otimes v$%
. Then the basic ingredient for constructing the XXX model is the following
simple solution to the Yang-Baxter equation%
\begin{equation}
r(\lambda )=\lambda +\tfrac{1}{2}+\tsum_{\alpha =1}^{3}\sigma ^{\alpha
}\otimes \sigma ^{\alpha }=\lambda +\mathbb{P}\in \limfunc{End}(\mathbb{C}%
^{2}\otimes \mathbb{C}^{2})  \label{r1}
\end{equation}%
Note that here we have changed our conventions from that in the introduction
as it simplifies some of the following computations. Another definition of
the XXX r-matrix is also commonly used in the literature,%
\begin{equation}
\tilde{r}(u):=ir(-iu-1/2)=u+i\tsum_{\alpha =1}^{3}\sigma ^{\alpha }\otimes
\sigma ^{\alpha }=u-i/2+i\mathbb{P\ }.  \label{r2}
\end{equation}%
Both definitions only differ by a re-parametrization of the spectral
parameter, $\lambda \rightarrow -iu-1/2,$ and an overall factor $i=\sqrt{-1}$%
. The functional relations and equations stated in the introduction refer to
this last convention (\ref{r2}).

In terms of (\ref{r1}) the transfer matrix of the inhomogeneous XXX model
with quasi-periodic boundary conditions is defined as follows,%
\begin{equation}
t_{\omega }(\lambda )=\limfunc{Tr}_{\mathbb{C}^{2}}\omega ^{\sigma
^{z}\otimes 1}r_{M}(\lambda -\lambda _{M})\cdots r_{1}(\lambda -\lambda
_{1})~,\qquad \omega =e^{i\phi }\ .
\end{equation}%
Here the trace is taken in the first factor of the $r$-matrix, i.e. $%
t_{\omega }\in \limfunc{End}(\mathbb{C}^{2})^{\otimes M}$. The set \{$%
\lambda _{m}$\}$_{m=1}^{M}$ are some arbitrary generic inhomogeneity
parameters, while the parameter $\omega =\exp (i\phi )$ incorporates the
twist angle $\phi $ which for the moment is allowed to be a generic complex
number, but can be specialized later on to real values in order to ensure
hermiticity. In the homogeneous limit $\lambda _{1}=...=\lambda _{M}=0$ its
meaning becomes apparent when writing down the associated spin-chain
Hamiltonian%
\begin{equation}
H_{\omega }=\left. \frac{d}{d\lambda }\ln \frac{t_{\omega }(\lambda )}{%
(\lambda +1)^{M}}\right\vert _{\lambda =0}=\frac{1}{2}\sum_{m=1}^{M}(\vec{%
\sigma}_{m}\cdot \vec{\sigma}_{m+1}-1)\ \;\;
\end{equation}%
with the boundary conditions%
\begin{equation}
\sigma _{M+1}^{x}\pm i\sigma _{M+1}^{y}=\omega ^{\pm 2}(\sigma _{1}^{x}\pm
i\sigma _{1}^{y})\qquad \text{and\qquad }\sigma _{M+1}^{z}=\sigma _{1}^{z}\ .
\end{equation}%
These boundary conditions break for $\omega \neq 1$ the spherical symmetry
of the Hamiltonian which unlike in the case of periodic boundary conditions
is not $sl_{2}$ invariant. However, there is an axial symmetry, i.e. the
total spin operator%
\begin{equation}
S^{z}=\frac{1}{2}\sum_{m=1}^{M}\sigma _{m}^{z}
\end{equation}%
is preserved. This breaking of the spherical symmetry is significant for the
Bethe ansatz analysis of the spectrum as for quasi-periodic boundary
conditions all eigenvectors become regular Bethe states. In the case of
periodic boundary conditions this is only true for the highest weight state
in each $sl_{2}$-module spanning one of the degenerate subspaces of the
transfer matrix respectively the Hamiltonian. This fact also plays an
important role in the construction of the $Q$-operator.

Besides the transfer matrix and the Hamiltonian it will be convenient to
discuss the entire fusion hierarchy of the $XXX$ model. To this end consider
the Chevalley-Serre generators of $sl_{2},$%
\begin{equation}
\lbrack h,e]=2e,\qquad \lbrack h,f]=-2f\qquad \text{and}\qquad \lbrack
e,f]=h,
\end{equation}%
then the following defines a well-known Verma module $\pi _{x}$ depending on
a complex parameter $x\in \mathbb{C}$,%
\begin{eqnarray}
\pi _{x}(e)\left\vert k\right\rangle &=&(x-k)k~\left\vert k-1\right\rangle
,\quad \;\pi _{x}(e)\left\vert 0\right\rangle =0  \label{pix} \\
\pi _{x}(f)\left\vert k\right\rangle &=&\left\vert k+1\right\rangle ,\quad 
\notag \\
\pi _{x}(h)\left\vert k\right\rangle &=&(x-2k-1)\left\vert k\right\rangle
,\qquad k=0,1,...,\infty \ .  \notag
\end{eqnarray}%
It is this Verma module which will form the auxiliary space for the $Q$%
-operator. Note that if $x=n\in \mathbb{N}_{>0}$ and one invokes the
truncation condition $\pi _{x}(f)\left\vert n\right\rangle =0,$ the $n$%
-dimensional subspace spanned by the vectors $\{\left\vert k\right\rangle
\}_{k=0}^{n-1}$ gives rise to the finite-dimensional modules $\pi ^{(n-1)}$
known as spin $s=(n-1)/2$ representations in the physics literature. Set%
\begin{equation}
L(\lambda )=\left( 
\begin{array}{cc}
\lambda +\frac{h+1}{2} & f \\ 
e & \lambda -\frac{h-1}{2}%
\end{array}%
\right) \in U(sl_{2})\otimes \limfunc{End}\mathbb{C}^{2}  \label{L}
\end{equation}%
then%
\begin{equation}
L_{12}(\lambda )L_{13}(\lambda +\lambda ^{\prime })r_{23}(\lambda ^{\prime
})=r_{23}(\lambda ^{\prime })L_{13}(\lambda +\lambda ^{\prime
})L_{12}(\lambda )
\end{equation}%
and the higher spin transfer matrix $t_{\omega }^{(n)}$ is defined through%
\begin{equation}
t_{\omega }^{(n)}(\lambda )=\limfunc{Tr}_{\pi ^{(n)}}\omega ^{h\otimes
1}L_{M}(\lambda -\lambda _{M})\cdots L_{1}(\lambda -\lambda _{1})\ .
\end{equation}%
The two distinguished elements in this hierarchy are the previously
introduced transfer matrix $t_{\omega }=t_{\omega }^{(1)}$ of spin $1/2$ and
the quantum determinant $\chi $ corresponding to the trivial representation
of spin $0,$%
\begin{equation}
\chi (\lambda )=t^{(0)}(\lambda )=\tprod_{m=1}^{M}(\lambda -\lambda _{m}+%
\tfrac{1}{2})\ .
\end{equation}%
From these two elements all the other members of the fusion hierarchy can be
generated via the the functional equation%
\begin{equation}
t_{\omega }^{(n)}(\lambda +\tfrac{n+1}{2})t_{\omega }^{(1)}(\lambda
)=t^{(0)}(\lambda +\tfrac{1}{2})t_{\omega }^{(n+1)}(\lambda +\tfrac{n}{2}%
)+t^{(0)}(\lambda -\tfrac{1}{2})t_{\omega }^{(n-1)}(\lambda +\tfrac{n+2}{2}%
)\ .  \label{fusion}
\end{equation}%
Instead of solving this functional relation in terms of $t_{\omega
}^{(1)},t^{(0)}$, which leads to quite involved formulae, it is simpler to
consider an auxiliary linear problem, Baxter's $TQ$ equation, which we
discuss next.

\section{The Q-operator and its spectrum}

We extend the definition of the higher spin transfer matrix to the
infinite-dimensional Verma module (\ref{pix}) introduced above and set%
\begin{equation}
Q_{\omega }(\lambda ;x)=\limfunc{Tr}_{\pi _{x}}\omega ^{h\otimes
1}L_{M}(\lambda -\lambda _{M}+\tfrac{x}{2})\cdots L_{1}(\lambda -\lambda
_{1}+\tfrac{x}{2})\ .
\end{equation}%
This definition of the $Q$-operator coincides with the isotropic limit of
the definition for the XXZ model \cite{CKQ7}. Note that the trace runs now
over an (half) infinite-dimensional space, whence it is crucial to have
quasi-periodic boundary conditions which upon the right choice of the twist
parameter $\omega $ ensure convergence \cite{CKQ3}.

Since the matrix $Q_{\omega }(\lambda ;x)$ preserves the total spin, $%
[Q_{\omega }(\lambda ;x),S^{z}]=0$, its matrix elements do always contain
the same number of the Chevalley-Serre generators $e$ and $f$. Using the
Casimir relation,%
\begin{equation}
\pi _{x}(C)=\frac{x^{2}-1}{2},\qquad C=h^{2}/2+h+2fe,
\end{equation}%
we deduce that it suffices to ensure that the\ following expressions are
finite%
\begin{equation}
\limfunc{Tr}_{\pi _{x}}\{\omega ^{h\otimes 1}h^{m}\}=\omega
^{x}\tsum_{k=0}^{\infty }\omega ^{-2k-1}(x-2k-1)^{m}<\infty ,\qquad
m=0,1,2,...,M\ .
\end{equation}%
This is obviously guaranteed as long as $|\omega |>1$. Employing the
geometric series to compute the trace, we then analytically continue this
operator from the region of convergence to the whole complex $\omega $%
-plane. Note that there remains a pole at $\omega =1$.

For instance, by construction $Q_{\omega }(\lambda ;x)$ is a polynomial of
degree $M$ in $\lambda $ and we have for the coefficient of the highest
power $\lambda ^{M},$%
\begin{equation}
Q_{\omega }(\lambda ;x)=\limfunc{Tr}_{\pi _{x}}\{\omega ^{h}\}~\lambda
^{M}+~...~=\sum_{k=0}^{\infty }\omega ^{x-2k-1}~\lambda ^{M}+~...~=\frac{%
\omega ^{x}}{\omega -\omega ^{-1}}~\lambda ^{M}+~...~,
\end{equation}%
where the last expression can be continued with respect to $\omega $ from
the region of convergence into the complex plane. Henceforth, this analytic
continuation from the region of convergence shall always be implicitly
understood.

The crucial property of the $Q$-operator is the following functional
equation, which strictly speaking is not yet Baxter's $TQ$ equation,%
\begin{equation}
t_{\omega }(\lambda )Q_{\omega }(\lambda ;x)=Q_{\omega }(\lambda
+1;x-1)\tprod_{m=1}^{M}(\lambda -\lambda _{m})+Q_{\omega }(\lambda
-1;x+1)\tprod_{m=1}^{M}(\lambda -\lambda _{m}+1)\ .
\end{equation}%
We omit the proof as it follows from taking the isotropic limit in the
analogous XXZ relations; see e.g. \cite{RW02,CKQ3,CKQ7}. The difference with
Baxter's $TQ$ equation is the fact that the additional complex parameter $x$
originating from the definition of the Verma module also shifts, instead of
only a shift in the spectral variable $\lambda $. Thus, the above equation
should rather be seen as an extension of the fusion hierarchy to
\textquotedblleft infinite\textquotedblright\ spin. Nevertheless, the
solutions to Baxter's TQ equation are obtained from $Q_{\omega }(\lambda ;x)$
through special limits. Namely, as we will discuss below we have the
following factorization%
\begin{equation}
Q_{\omega }(\lambda ;x)=\frac{\omega ^{x}}{\omega -\omega ^{-1}}Q_{\omega
}^{+}(\lambda )Q_{\omega }^{-}(\lambda +x),  \label{Qfactor}
\end{equation}%
where $Q_{\omega }^{\pm }$ are two linearly independent solutions to
Baxter's $TQ$ equation%
\begin{equation}
t_{\omega }(\lambda )Q_{\omega }^{\pm }(\lambda )=\omega ^{\mp 1}Q_{\omega
}^{\pm }(\lambda +1)\tprod_{m=1}^{M}(\lambda -\lambda _{m})+\omega ^{\pm
1}Q_{\omega }^{\pm }(\lambda -1)\tprod_{m=1}^{M}(\lambda -\lambda _{m}+1)\ .
\label{twqBx}
\end{equation}%
We now turn to the discussion of the spectrum of the Q operator where we
will explain in more detail the above factorization into the solutions $%
Q_{\omega }^{\pm }$.

\subsection{The algebraic Bethe ansatz analysis of $Q$}

In the context of the XXZ model the spectrum of the $Q$-operator has been
analyzed \cite{CKQ3} using the formalism of the algebraic Bethe ansatz \cite%
{FST79}. We recall that for twisted boundary conditions there is no problem
with the Bethe ansatz as the $sl_{2}$ symmetry of the XXX model is broken
and all eigenstates of the XXX transfer matrix are regular Bethe states; see
for instance the discussion in \cite{TV95} where the completeness of the
Bethe ansatz in a neighbourhood of $\omega =0$ has been shown. Using the
analogous algebraic relations as in the XXZ case \cite{CKQ3}, one can show
that the Bethe states are eigenvectors of the $Q$-operator\footnote{%
At the moment this has only been carried out for Bethe states with $n<4$ due
to the complicated and numerous unwanted terms, see the appendix in \cite%
{CKQ3}. However, in the case of the XXZ model alternative proofs (based on
functional relations) exist \cite{CKQ2,CKQ4} which match the algebraic Bethe
ansatz result for arbitrary $n$. The spectrum for the XXX model presented
here is the isotropic limit of the XXZ result \cite{CKQ3,CKQ7}.}. Namely,
decomposing the monodromy matrix of the XXX model in the usual manner%
\begin{equation}
\boldsymbol{t}(\lambda )=\omega ^{\sigma ^{z}\otimes 1}r_{M}(\lambda
-\lambda _{M})\cdots r_{1}(\lambda -\lambda _{1})=\left( 
\begin{array}{cc}
A(\lambda ) & B(\lambda ) \\ 
C(\lambda ) & D(\lambda )%
\end{array}%
\right) 
\end{equation}%
one considers for $n=M/2-S^{z}>0$ an \textquotedblleft
admissible\textquotedblright\ solution \cite{TV95} to the Bethe ansatz
equations above the equator (note that according to the conventions (\ref{r1}%
) and (\ref{r2}) the corresponding Bethe roots are related by $\xi
_{j}^{+}\rightarrow -iv_{j}^{+}-1/2$) 
\begin{subequations}
\begin{equation}
\omega ^{-1}\tprod_{j=1}^{n}(\xi _{i}^{+}-\xi
_{j}^{+}+1)\tprod_{m=1}^{M}(\xi _{i}^{+}-\lambda _{m})+\omega
\tprod_{j=1}^{n}(\xi _{i}^{+}-\xi _{j}^{+}-1)\tprod_{m=1}^{M}(\xi
_{i}^{+}-\lambda _{m}+1)=0\ .  \label{bae2}
\end{equation}%
Then it follows from the Yang-Baxter equation that the matrix elements $%
\{Q_{kl}\}_{k,l\in \mathbb{N}}$ of the monodromy matrix 
\end{subequations}
\begin{equation*}
\boldsymbol{Q}(\lambda )=\omega ^{h\otimes 1}L_{M}(\lambda -\lambda _{M}+%
\tfrac{x}{2})\cdots L_{1}(\lambda -\lambda _{1}+\tfrac{x}{2})
\end{equation*}%
with respect to the infinite-dimensional auxiliary space corresponding to $%
\pi _{x}$ satisfy certain commutation relations with the Yang-Baxter algebra 
$\{A,B,C,D\}$, for example \cite{CKQ3}%
\begin{multline*}
Q_{k,l}(\lambda )B(\xi )=\frac{\alpha _{l+1}\delta _{l}-\beta _{l+1}\gamma
_{l}}{\alpha _{k}\alpha _{l+1}}~B(\xi )Q_{k,l}(\lambda ) \\
+\frac{\beta _{l+1}}{\alpha _{l+1}}~Q_{k,l+1}(\lambda )A(\xi )-\frac{\beta
_{k}}{\alpha _{k}}~Q_{k+1,l}(\lambda )D(\xi )+\frac{\beta _{k}\beta _{l+1}}{%
\alpha _{k}\alpha _{l+1}}Q_{k+1,l+1}(\lambda )C(\xi ),
\end{multline*}%
where the coefficients are determined through the matrix elements of the $L$%
-operator (\ref{L}),%
\begin{equation}
\alpha _{k}=\lambda -\xi +x-k,\;\;\beta _{k}=1,\;\;\gamma
_{k-1}=(x-k)k,\;\;\delta _{k}=\lambda -\xi +k+1\;.  \label{coeff}
\end{equation}%
For a complete list of the algebraic identities we refer the reader to \cite%
{CKQ3}. Employing these commutation relations one can identify the
eigenvalue of the $Q$-operator on a Bethe state. Denoting by $\left\vert
0\right\rangle $ the pseudo vacuum, i.e. the state with all spins up, the
Bethe vector associated with an admissible solution to (\ref{bae2}) is an
eigenstate of the $Q$-operator with eigenvalue 
\begin{equation}
Q_{\omega }(\lambda ;x)B(\xi _{1}^{+})\cdots B(\xi _{n}^{+})\left\vert
0\right\rangle =\frac{\omega ^{x}}{\omega -\omega ^{-1}}Q_{\omega
}^{+}(\lambda )Q_{\omega }^{-}(\lambda +x)B(\xi _{1}^{+})\cdots B(\xi
_{n}^{+})\left\vert 0\right\rangle 
\end{equation}%
where%
\begin{equation}
Q_{\omega }^{+}(\lambda )=\tprod_{j=1}^{n}(\lambda -\xi _{j}^{+})  \label{Qp}
\end{equation}%
and%
\begin{equation}
Q_{\omega }^{-}(\lambda )=\left( \omega -\omega ^{-1}\right) Q_{\omega
}^{+}(\lambda )\sum_{k=0}^{\infty }\frac{\omega ^{-2k-1}\prod_{m}(\lambda
-\lambda _{m}-k)}{Q_{\omega }^{+}(\lambda -k)Q_{\omega }^{+}(\lambda -k-1)}\
.
\end{equation}%
Notice that $Q_{\omega }^{-}$ is polynomial in $\lambda $ due to the the
Bethe ansatz equations. In fact, by the very construction of the $Q$%
-operator it must be a polynomial of degree $M-n$,%
\begin{equation}
Q_{\omega }^{-}(\lambda )=\tprod_{j=1}^{M-n}(\lambda -\xi _{j}^{-})\ .
\label{Qm}
\end{equation}%
Exploiting the completeness of the Bethe ansatz for generic quasi-periodic
boundary conditions and inhomogeneity parameters \cite{TV95}, we obtain the
factorization of the $Q$-operator into the previously introduced, linearly
independent solutions $Q_{\omega }^{\pm }$ of Baxter's $TQ$ equation (\ref%
{twqBx}). We might define them implicitly as operators through the following
limits%
\begin{equation}
\lim_{x\rightarrow -\lambda }Q_{\omega }(\lambda ;x)=\frac{\omega ^{-\lambda
}}{\omega -\omega ^{-1}}Q_{\omega }^{+}(\lambda )Q_{\omega }^{-}(0)
\end{equation}%
and%
\begin{equation}
\lim_{\lambda \rightarrow 0}Q_{\omega }(\lambda ;x)=\frac{\omega ^{x}}{%
\omega -\omega ^{-1}}Q_{\omega }^{+}(0)Q_{\omega }^{-}(x)\ .
\end{equation}%
We shall denote the operators and eigenvalues by the same symbol. In
contrast to the XXZ case \cite{CKQ7}\ the operators $Q_{\omega }^{\pm }(0)$
are not easily determined and we are missing at the moment concrete operator
expressions for them. However, explicit computation of the $Q$-operators in
the various spin-sectors for small lattice sizes ($M\leq 6$) shows that the
following expressions drastically simplify%
\begin{equation}
\omega ^{\lambda }Q_{\omega }(\lambda ;-\lambda )Q_{\omega
}(0;0)^{-1}=Q_{\omega }^{+}(\lambda )Q_{\omega }^{+}(0)^{-1}  \label{Qp2}
\end{equation}%
and%
\begin{equation}
\omega ^{-\lambda }Q_{\omega }(0;0)^{-1}Q_{\omega }(0;\lambda )=Q_{\omega
}^{-}(0)^{-1}Q_{\omega }^{-}(\lambda )\ .  \label{Qm2}
\end{equation}%
Both (\ref{Qp}) and (\ref{Qm}) are obviously solutions to Baxter's TQ
equation (\ref{twqBx}) and are normalized to the identity matrix at the
origin $\lambda =0$. The inverse matrices exist as long as none of the Bethe
roots $\xi _{j}^{\pm }$ vanishes, which is the case as long as $\omega \neq 1
$. Despite this lack of information on the normalization constants, our $Q$%
-operator analysis yields computational advantages. Before we address the
latter let us first discuss the relationship between $Q_{\omega }^{\pm
}(\lambda )$ under spin reversal.

\subsection{Spin reversal}

Define the spin reversal operator by setting $\mathfrak{R}%
=\prod_{m=1}^{M}\sigma _{m}^{x}$. Due to the twisted boundary conditions
spin reversal symmetry is broken and we have for the transfer matrix the
identity%
\begin{equation}
\mathfrak{R~}t_{\omega }(\lambda )~\mathfrak{R}=t_{\omega ^{-1}}(\lambda )\ .
\end{equation}%
Let us now investigate the transformation of the $Q$-operator under spin
reversal. From the equality%
\begin{equation}
(1\otimes \sigma ^{x})L(\lambda )(1\otimes \sigma ^{x})=-\left( 
\begin{array}{cc}
-\lambda -1+\frac{h+1}{2} & -e \\ 
-f & -\lambda -1-\frac{h-1}{2}%
\end{array}%
\right)
\end{equation}%
it follows for the homogeneous model $\lambda _{1}=...=\lambda _{M}=0$ that%
\begin{equation}
\mathfrak{R}~Q_{\omega }(\lambda ;x)~\mathfrak{R}=(-)^{M}Q_{\omega
}(-\lambda -1-x;x)^{t}\ .  \label{RQR1}
\end{equation}%
Alternatively, we can compute the spectrum of $\breve{Q}_{\omega }:=%
\mathfrak{R}Q_{\omega }\mathfrak{R}$ from the algebraic Bethe ansatz similar
as before. In terms of the matrix elements of the associated monodromy
matrices the basic relation we need is 
\begin{equation}
\breve{Q}_{k,l}B=\left( \frac{\alpha _{l}}{\delta _{k}}-\frac{\gamma
_{l-1}\beta _{l}}{\delta _{k}\delta _{l-1}}\right) B\breve{Q}_{k,l}+\frac{%
\gamma _{l-1}}{\delta _{l-1}}~\breve{Q}_{k,l-1}A-\frac{\gamma _{k}}{\delta
_{k}}~\breve{Q}_{k+1,l}D+\frac{\gamma _{k}\gamma _{l-1}}{\delta _{k}\delta
_{l-1}}\breve{Q}_{k+1,l-1}C\ .
\end{equation}%
Here the coefficients are the same as in (\ref{coeff}). This then leads to
the following eigenvalues corresponding to Bethe states%
\begin{multline*}
\mathfrak{R}Q_{\omega }(\lambda ;x)\mathfrak{R}B(\xi _{1}^{+})\cdots B(\xi
_{n}^{+})\left\vert 0\right\rangle = \\
Q_{\omega ^{-1}}^{+}(\lambda +x)Q_{\omega ^{-1}}^{+}(\lambda
)\sum_{k=0}^{\infty }\frac{\omega ^{x-2k-1}\prod_{m}(\lambda -\lambda
_{m}+k+1)}{Q_{\omega ^{-1}}^{+}(\lambda +k)Q_{\omega ^{-1}}^{+}(\lambda +k+1)%
}~B(\xi _{1}^{+})\cdots B(\xi _{n}^{+})\left\vert 0\right\rangle
\end{multline*}%
As already previously mentioned for generic inhomogeneity parameters $%
\lambda _{m}$ and a suitable neighbourhood of $\omega =0$ (or $\omega
=\infty $) the Bethe ansatz yields a complete set of eigenstates \cite{TV95}%
. This fact now implies the operator equation%
\begin{equation}
\mathfrak{R}Q_{\omega }(\lambda ;x)\mathfrak{R}=Q_{\omega ^{-1}}(\lambda
+x;-x)=-\frac{\omega ^{x}}{\omega -\omega ^{-1}}~Q_{\omega
^{-1}}^{+}(\lambda +x)Q_{\omega ^{-1}}^{-}(\lambda )\ .  \label{RQR2}
\end{equation}%
Therefore, under spin reversal the roles of $Q_{\omega }^{+},Q_{\omega }^{-}$
are interchanged. These relations match the analogous ones derived for the
six-vertex model \cite{CKQ3,CKQ4,CKQ7}.

\section{Fusion hierarchy and complex dimension}

One of the aforementioned advantages of our $Q$-operator analysis is that
the relation between $Q_{\omega }(\lambda ;x)$ and the higher spin transfer
matrices $t_{\omega }^{(n-1)}$ is particularly simple allowing one through
analytic continuation to compactly present the information on the entire
fusion hierarchy. Specializing $x\rightarrow n\in \mathbb{N},$ it was
already pointed out earlier that the infinite-dimensional Verma module (\ref%
{pix}) contains a finite dimensional subrepresentation spanned by the
vectors $\{\left\vert k\right\rangle \}_{k=0}^{n-1}$ and which is isomorphic
to the $sl_{2}$ representation $\pi ^{(n-1)}$ of spin $s=(n-1)/2$. The
remaining space spanned by $\{\left\vert k\right\rangle \}_{k=n}^{\infty }$
can be identified again as the Verma module $\pi _{x}$ with $x=-n$. This
simple representation theoretic fact translates into the following
functional relation when splitting the trace over the aforementioned
subspaces,%
\begin{equation}
t_{\omega }^{(n-1)}(\lambda )=Q_{\omega }(\lambda -\tfrac{n}{2};n)-Q_{\omega
}(\lambda +\tfrac{n}{2};-n)\ .  \label{twn}
\end{equation}%
Thus the spectrum of the higher spin transfer matrices takes a particularly
simple form in terms of the spectrum of $Q_{\omega }(\lambda ;x)$. In
contrast the expression from the algebraic Bethe ansatz and the fusion
relation (\ref{fusion}) is more involved. Furthermore, we might analytically
continue expression (\ref{twn}) in the spin variable $n$ setting%
\begin{equation}
t_{\omega }(\lambda ;x)=Q_{\omega }(\lambda -\tfrac{x}{2};x)-Q_{\omega
}(\lambda +\tfrac{x}{2};-x)\ .  \label{twx}
\end{equation}%
The last object combines the information of the entire fusion hierarchy.
Notice that in (\ref{twn}) respectively (\ref{twx}) one can safely take the
limit to periodic boundary conditions, i.e. the following object is well
defined%
\begin{equation}
t(\lambda ;x)=\lim_{\omega \rightarrow 1}t_{\omega }(\lambda
;x)=\lim_{\omega \rightarrow 1}\left[ Q_{\omega }(\lambda -\tfrac{x}{2}%
;x)-Q_{\omega }(\lambda +\tfrac{x}{2};-x)\right] \ .  \label{tx}
\end{equation}%
In this manner one recovers the XXX model with periodic boundary conditions.
The transfer matrix $t(\lambda ;x)$ with \textquotedblleft complex
dimension\textquotedblright\ $x$ coincides with the generalized trace
construction \cite{BJMSTxxx} in the context of correlation functions on the
infinite lattice. This complex dimension occurs as the coefficient of the
highest power in the polynomial $t(\lambda ;x)$,%
\begin{equation}
t(\lambda ;x)=x\lambda ^{M}+\sum_{m=0}^{M-1}t_{m}(x)\lambda ^{m}\ .
\end{equation}%
In comparison, the analogous result in the context of the six-vertex or XXZ
model showed the appearance of logarithmic terms; see \cite{CKQ7}.

\subsection{The trace functional: a simple example $M=4,\ S^{z}=0$}

It is instructive to verify for a simple example whether the construction (%
\ref{tx}) coincides with the definition through the trace functional given
in \cite{BJMSTxxx}. Setting $M=4$ and $S^{z}=0$ we consider a diagonal
matrix element of the $Q$-operator,%
\begin{equation}
Q_{\omega }(\lambda ;x)_{\alpha _{1}...\alpha _{4}}^{\alpha _{1}...\alpha
_{4}}=\sum_{k=0}^{\infty }\omega ^{x-2k-1}(\lambda +x-k)^{2}(\lambda
+k+1)^{2}
\end{equation}%
Here $\alpha _{i}=\pm 1$ are the eigenvalues of $\sigma _{i}^{z}$ acting on
the $i^{\text{th}}$ lattice site with $i=1,2,3,4$ and $\sum_{i}\alpha _{i}=0$%
. Using the formula for the geometric series and analytically continuing the
result in $\omega $ afterwards to take the limit $\omega \rightarrow 1$ in (%
\ref{tx}) we arrive at%
\begin{equation}
t(\lambda ;x)_{\alpha _{1}...\alpha _{4}}^{\alpha _{1}...\alpha _{4}}=\frac{%
32x-20x^{3}+3x^{5}}{240}+\frac{4x-x^{3}}{6}~\lambda +\frac{10x-x^{3}}{6}%
~\lambda ^{2}+2x~\lambda ^{3}+x~\lambda ^{4}\ .
\end{equation}%
The action of the trace functional $\limfunc{Tr}_{x}:U(sl_{2})\otimes 
\mathbb{C}[x]\rightarrow \mathbb{C}[x]$ introduced in \cite{BJMSTxxx} (not
to be mistaken for $\limfunc{Tr}_{\pi _{x}}\neq \limfunc{Tr}_{x}$) on the
powers of the Cartan generators is defined through 
\begin{equation}
\limfunc{Tr}_{x}\{e^{zh}\}=\frac{\sinh (zx)}{\sinh z}=x+\frac{x^{3}-x}{6}%
~z^{2}+\frac{7x-10x^{3}+3x^{5}}{360}~z^{4}+~...  \label{trxh}
\end{equation}%
Acting now with the trace functional on the monodromy matrix of $L$%
-operators we compute%
\begin{multline*}
\limfunc{Tr}_{x}L(\lambda )_{\alpha _{4}}^{\alpha _{4}}L(\lambda )_{\alpha
_{3}}^{\alpha _{3}}L(\lambda )_{\alpha _{2}}^{\alpha _{2}}L(\lambda
)_{\alpha _{1}}^{\alpha _{1}}=\limfunc{Tr}_{x}\{(\lambda +\tfrac{h+1}{2}%
)^{2}(\lambda -\tfrac{h-1}{2})^{2}\}= \\
\frac{\limfunc{Tr}_{x}\{1-2h^{2}+h^{4}\}}{16}+\frac{\limfunc{Tr}%
_{x}\{1-h^{2}\}}{2}~\lambda +\frac{\limfunc{Tr}_{x}\{3-h^{2}\}}{2}~\lambda
^{2}+2\limfunc{Tr}_{x}\{1\}~\lambda ^{3}+\limfunc{Tr}_{x}\{1\}~\lambda ^{4}=
\\
t(\lambda ;x)_{\alpha _{1}...\alpha _{4}}^{\alpha _{1}...\alpha _{4}},
\end{multline*}%
where the last line is obtained after inserting the values from the
expansion (\ref{trxh}). Thus, we find agreement with (\ref{tx}). To
illustrate the generalized transfer matrix of complex dimension further we
present its eigenvalues in the table below. Specializing $x$ to be an
integer \TEXTsymbol{>} 0 one obtains the eigenvalues of each element in the
fusion hierarchy.\medskip 

\begin{center}
\begin{tabular}{|c|c|}
\hline\hline
$P$ & $t(\lambda ;x)$ \\ \hline\hline
$\pi $ & $\frac{x}{2}-\frac{x^{3}}{2}+\frac{x^{5}}{16}+\frac{2x-x^{3}}{2}%
\,\lambda +\frac{4x-x^{3}}{2}\,{\lambda }^{2}+2\,x\,{\lambda }^{3}+x\,{%
\lambda }^{4}$ \\ \hline\hline
$\pi $ & $\frac{x}{6}-\frac{x^{3}}{12}-\frac{x^{5}}{48}+\frac{4x-x^{3}}{6}%
\,\lambda +\frac{10x-x^{3}}{6}\,{\lambda }^{2}+2\,x\,{\lambda }^{3}+x\,{%
\lambda }^{4}$ \\ \hline\hline
$0$ & $\frac{x}{6}-\frac{x^{3}}{6}+\frac{x^{5}}{16}+\frac{2x-x^{3}}{2}%
\,\lambda +\frac{4x-x^{3}}{2}\,{\lambda }^{2}+2\,x\,{\lambda }^{3}+x\,{%
\lambda }^{4}$ \\ \hline\hline
$0$ & $-\frac{x}{30}+\frac{x^{3}}{12}+\frac{x^{5}}{80}+\frac{x^{3}\,\lambda 
}{2}+\frac{2x+x^{3}}{2}\,{\lambda }^{2}+2\,x\,{\lambda }^{3}+x\,{\lambda }%
^{4}$ \\ \hline\hline
$\pi /2$ & $\frac{8ix+(4-8i)x^{3}-x^{5}}{48}+\frac{(4+2i)x-(1+2i)x^{3}}{6}%
\,\lambda +\frac{10x-x^{3}}{6}\,{\lambda }^{2}+2\,x\,{\lambda }^{3}+x\,{%
\lambda }^{4}$ \\ \hline\hline
$\pi /2$ & $-\frac{8ix-(4+8i)x^{3}+x^{5}}{48}+\frac{(4-2i)x-(1-2i)x^{3}}{6}%
\,\lambda +\frac{10x-x^{3}}{6}\,{\lambda }^{2}+2\,x\,{\lambda }^{3}+x\,{%
\lambda }^{4}$ \\ \hline\hline
\end{tabular}%
\medskip

{\small Table 1. Spectrum of the transfer matrix with complex dimension 
\emph{x}.}
\end{center}

\section{The quantum Wronskian}

The second computational advantage from the $Q$-operator analysis is of
great practical importance in the actual computation of the spectra of the
Hamiltonian and the transfer matrices. Instead of solving the quite
intricate Bethe ansatz equations, one can now turn the ideology around and
rather interpret the relation (\ref{twx}) for $x=1$, named the quantum
Wronskian, as the fundamental identity,%
\begin{equation}
\tprod_{m=1}^{M}(\lambda -\lambda _{m})=\frac{\omega ~Q_{\omega
}^{+}(\lambda -1)Q_{\omega }^{-}(\lambda )-\omega ^{-1}Q_{\omega
}^{+}(\lambda )Q_{\omega }^{-}(\lambda -1)}{\omega -\omega ^{-1}}\ .
\label{Wronski}
\end{equation}%
Here we have exploited the factorization (\ref{Qfactor}). In terms of the
eigenvalues (\ref{Qp}), (\ref{Qm}) the above relation incorporates the Bethe
ansatz equations above and below the equator with respect to the
parametrization (\ref{r1}),%
\begin{equation}
\tprod_{m=1}^{M}\frac{\xi _{i}^{\pm }-\lambda _{m}}{\xi _{i}^{\pm }-\lambda
_{m}+1}=\omega ^{\pm 2}\tprod_{j=1}^{n_{\pm }}\frac{\xi _{i}^{\pm }-\xi
_{j}^{\pm }-1}{\xi _{i}^{\pm }-\xi _{j}^{\pm }+1},\qquad n_{\pm }=M/2\mp
S^{z},  \label{BAExxx}
\end{equation}%
and is therefore sufficient to analyze the spectrum. Introducing the
elementary symmetric polynomials $e_{k}^{\pm }=e_{k}(\xi _{1}^{\pm },...,\xi
_{n_{\pm }}^{\pm })$ in the Bethe roots%
\begin{equation}
Q_{\omega }^{\pm }(\lambda )=\sum_{k=0}^{n_{\pm }}(-)^{k}e_{k}^{\pm
}~\lambda ^{n_{\pm }-k},
\end{equation}%
the quantum Wronskian (\ref{Wronski}) becomes the following identity%
\begin{equation}
e_{M-m}(\lambda _{1},...,\lambda _{M})=\sum_{k=0}^{m}\sum_{\ell \geq m-k}%
\binom{\ell }{m-k}\frac{\omega ~e_{n-\ell }^{+}e_{M-n-k}^{-}-\omega
^{-1}e_{n-k}^{+}e_{M-n-\ell }^{-}}{\omega -\omega ^{-1}},
\end{equation}%
which is quadratic in the $M$ unknowns $e_{k}^{\pm }$. Here $e_{m}(\lambda
_{1},...,\lambda _{M})$ is the $m^{\text{th}}$ elementary symmetric
polynomial in the inhomogeneity parameters. Furthermore, we use the
convention $e_{k}^{\pm }\equiv 0$ for $k<0$ and $k>n_{\pm }=M/2\mp S^{z}$.
In contrast the Bethe ansatz equations (\ref{BAExxx}) are of order $M$. The
approach based on the quantum Wronskian (\ref{Wronski}) therefore leads to a
significant advantage in numerical computations for long spin-chains.

Note that in the limits $\omega \rightarrow 0,$~$\infty $ we can easily
establish the completeness of the Bethe ansatz for generic inhomogeneity
parameters by a similar line of argument as it has been used in \cite{TV95}.
Namely, assuming all inhomogeneity parameters $\{\lambda _{j}\}$ to be
mutually distinct we infer from the quantum Wronskian (\ref{Wronski}) the
solutions%
\begin{equation*}
\omega =\infty :\quad Q_{\infty }^{+}(\lambda )=\tprod_{j=1}^{n}(\lambda
-\lambda _{m_{j}}+1)\qquad \text{and\qquad }Q_{\infty }^{-}(\lambda
)=\tprod_{j=1}^{M-n}(\lambda -\lambda _{m_{j+n}})
\end{equation*}%
for any permutation $(m_{1},...,m_{M})$ of the index set $\{1,2,...,M\}$.
Obviously, the number of distinct solutions is then $\binom{M}{n}$ which
coincides with the dimension of the associated spin sector. For $\omega =0$
the roles of $Q^{\pm }$ are interchanged. Using the implicit function
theorem one can then argue that the number of solutions stays the same in
the vicinity of the point $\omega =\infty $ respectively $\omega =0$.

\subsection{Special solutions for homogeneous chains of even length and $%
S^{z}=0$}

Let $M\in 2\mathbb{N}$ and consider the spin sector $S^{z}=0$. Then
according to our previous discussion $Q_{\omega }^{+}$ and $Q_{\omega }^{-}$
have the same polynomial degree $n=M/2$ and in light of (\ref{RQR1}), (\ref%
{RQR2}) one might expect a simple relationship between them. In fact, based
on numerical studies of homogeneous spin-chains up to length $M=10$ and $%
\omega =e^{i\phi },\ \phi \in \mathbb{R}$ one confirms that there exist $%
2^{M/2}$ solutions to the quantum Wronskian which satisfy%
\begin{equation}
M\in 2\mathbb{N},\;S^{z}=0:\quad \quad Q_{\omega }^{-}(\lambda )=(-1)^{\frac{%
M}{2}}Q_{\omega }^{+}(-\lambda -1)\;.  \label{specialwronski}
\end{equation}%
Notably, for the mentioned examples $M=2,4,6,8,10$ the eigenvalue of the
transfer matrix which belongs to the groundstate in the limit $\omega
\rightarrow 1$ always appears to be among this set of special solutions.

For the numerical investigation it is more convenient to use the second
parametrization (\ref{r2}) of the XXX model, since then the coefficients
(not the roots) of the polynomials%
\begin{equation*}
\tilde{Q}_{\omega }^{\pm }(u)=i^{n_{\pm }}Q_{\omega }^{\pm
}(-iu-1/2)=\tprod_{j=1}^{n_{\pm }}(u-v_{j}^{\pm })
\end{equation*}%
are always real numbers. In this parametrization the special relationship (%
\ref{specialwronski}) becomes simply%
\begin{equation}
M\in 2\mathbb{N},\;S^{z}=0:\quad \quad \tilde{Q}_{\omega }^{-}(u)=(-1)^{%
\frac{M}{2}}\tilde{Q}_{\omega }^{+}(-u)\ .  \label{specialw2}
\end{equation}%
At the moment there is no derivation from first principles for this
simplification, however, it can be motivated by (\ref{RQR1}) which states
that left and right eigenvectors of the Q-operator are related by
spin-reversal. As the spin zero sector is invariant under the action of the
spin-reversal operator it can happen that some left and right eigenvectors
of Q coincide leading via (\ref{RQR1}) to the simplification (\ref%
{specialwronski}) respectively (\ref{specialw2}). Assuming the latter to
hold true one can verify it for chains of length $M>10$ by inserting this
special subset of solutions into the Wronskian relation which then
simplifies to%
\begin{equation}
u^{M}=(-1)^{\frac{M}{2}}\frac{\omega \tilde{Q}_{\omega }^{+}(u-\frac{i}{2})%
\tilde{Q}_{\omega }^{+}(-u-\frac{i}{2})-\omega ^{-1}\tilde{Q}_{\omega
}^{+}(u+\frac{i}{2})\tilde{Q}_{\omega }^{+}(-u+\frac{i}{2})}{\omega -\omega
^{-1}}\ .
\end{equation}%
Specializing the spectral parameter to $u=v_{j}^{+}+\frac{i}{2}$ and $%
u=v_{j}^{+}-\frac{i}{2}$ we now obtain the following sets of equations for
the Bethe roots $v_{j}^{+}$ of this subclass of solutions%
\begin{equation}
(v_{j}^{+}+i/2)^{M}=\frac{\omega ^{-1}}{\omega ^{-1}-\omega }%
~\tprod_{k=1}^{M/2}(v_{j}^{+}-v_{k}^{+}+i)(v_{j}^{+}+v_{k}^{+})
\label{special1}
\end{equation}%
and%
\begin{equation}
(v_{j}^{+}-i/2)^{M}=\frac{\omega }{\omega -\omega ^{-1}}%
~\tprod_{k=1}^{M/2}(v_{j}^{+}-v_{k}^{+}-i)(v_{j}^{+}+v_{k}^{+}),
\label{special2}
\end{equation}%
respectively. Since $\omega $ lies on the unit circle both equations are
equivalent under complex conjugation provided the Bethe roots $v_{j}^{+}$
are either real or occur in complex conjugate pairs. For the mentioned
examples this is indeed the case. Dividing these two equations yields the
familiar Bethe ansatz equations for twisted boundary conditions,%
\begin{equation}
\left( \frac{v_{j}^{+}+i/2}{v_{j}^{+}-i/2}\right) ^{M}=\omega
^{-2}\prod_{k\neq j}^{M/2}\frac{v_{j}^{+}-v_{k}^{+}+i}{v_{j}^{+}-v_{k}^{+}-i}%
\ .
\end{equation}%
Thus, we infer that extending the assumption (\ref{specialw2}) beyond the
numerically checked examples of spin-chains of length $M\leq 10$ is
compatible with the Bethe ansatz. The corresponding eigenvalues of the
transfer matrix are of the form%
\begin{equation}
\tilde{t}(u)=i^{M}t(-iu-1/2)=(\omega +\omega ^{-1})u^{M}+\sum_{m=1}^{M/2}%
\tilde{t}_{m}u^{M-2m},
\end{equation}%
i.e. only even powers of the spectral parameter $u$ occur. In addition, the
parameters $\tilde{t}_{m}$ are real and the eigenvalue corresponding to the
groundstate in the limit of periodic boundary conditions $\omega \rightarrow
1$ is distinguished by the fact that all coefficients have the same sign, $%
\limfunc{sgn}\tilde{t}_{m}=\limfunc{sgn}(\omega +\omega ^{-1})$. We leave a
more detailed study of these solutions to future work as it involves more
extensive numerical calculations.\medskip 

\begin{center}
\hspace{-0.5cm}%
\begin{tabular}{|l|l|}
\hline\hline
$\phi $ & $M=10$ \\ \hline\hline
$\dfrac{\pi }{2}$ & 
\begin{tabular}{l}
${\small Q}^{+}{\small =u}^{5}{\small \mp 0.7769661~u}^{4}{\small %
-0.3231618~u}^{3}{\small \pm 0.1117312~u}^{2}{\small +0.011890969~u\mp
0.01189097}$ \\ 
${\small t=\pm 1.553932~u}^{8}{\small \pm 6.04751~u}^{6}{\small \pm 9.74055~u%
}^{4}{\small \pm 7.58483~u}^{2}{\small \pm 2.37584}$%
\end{tabular}
\\ \hline\hline
$\dfrac{\pi }{20}$ & 
\begin{tabular}{l}
${\small Q}^{+}{\small =u}^{5}{\small -0.06935158~u}^{4}{\small -0.403661~u}%
^{3}{\small +0.01057672~u}^{2}{\small -0.0166721~u-0.000107935}$ \\ 
${\small t=1.97538~u}^{10}{\small +7.42936~u}^{8}{\small +13.5893~u}^{6}%
{\small +14.8551~u}^{4}{\small +9.33335~u}^{2}{\small +2.59013}$%
\end{tabular}
\\ \hline\hline
$\dfrac{\pi }{200}$ & 
\begin{tabular}{l}
${\small Q}^{+}{\small =u}^{5}{\small -0.00692881~u}^{4}{\small -0.404443~u}%
^{3}{\small +0.00105731~u}^{2}{\small -0.0167203~u-0.000107938}$ \\ 
${\small t=1.99975~u}^{10}{\small +7.49929~u}^{8}{\small +13.6758~u}^{6}%
{\small +14.9119~u}^{4}{\small +9.35224~u}^{2}{\small +2.59241}$%
\end{tabular}
\\ \hline\hline
$0$ & 
\begin{tabular}{l}
${\small Q}^{+}{\small =u}^{5}{\small -0.404451~u}^{3}{\small -0.0167203~u}$
\\ 
${\small t=2u}^{10}{\small +}\frac{15}{2}{\small ~u}^{8}{\small +13.6767~u}%
^{6}{\small +14.9125~u}^{4}{\small +9.35243~u}^{2}{\small +2.59243}$%
\end{tabular}
\\ \hline\hline
\end{tabular}%
\medskip
\end{center}

\noindent {\small Table 2. Groundstate eigenvalues of the transfer matrix
and Q-operator in the spin zero sector for various twist parameters.}

\subsection{Eigenvalues in the limit of periodic boundary conditions}

Let us make contact with the discussion of Pronko and Stroganov for the XXX
model with periodic boundary conditions $\phi =0$ respectively $\omega =1$ 
\cite{PS99}. Starting from the $TQ$ equation on the level of eigenvalues
they reported the following quantum Wronskian relation with respect to the
parametrization (\ref{r2}),%
\begin{equation}
\mathcal{Q}^{-}(u+\tfrac{i}{2})\mathcal{Q}^{+}(u-\tfrac{i}{2})-\mathcal{Q}%
^{-}(u-\tfrac{i}{2})\mathcal{Q}^{+}(u+\tfrac{i}{2})=u^{M}  \label{PSwronski}
\end{equation}%
with the crucial difference that the degree of the second linearly
independent solution $\mathcal{Q}^{-}$ is now increased by one,%
\begin{equation}
\mathcal{Q}^{-}(u)=\frac{-i}{2S^{z}+1}\prod_{k=1}^{M-n+1}(u-v_{k}^{-})\ .
\end{equation}%
The degree of the other solution, $\mathcal{Q}^{+},$ describing the well
known Bethe roots above the equator remains unchanged,%
\begin{equation}
\mathcal{Q}^{+}(u)=\tprod_{k=1}^{n}(u-v_{k}^{+})\ .
\end{equation}%
We have deliberately denoted their solutions $\mathcal{Q}^{\pm }$ by a
different symbol to distinguish them from the solutions $Q_{\omega }^{\pm }$
obtained from our operator construction at quasi-periodic boundary
conditions. As already pointed out in the introduction the quantum Wronskian
(\ref{PSwronski}) has a restricted number of solutions which is much smaller
than the dimension of the respective spin sector fixed by the degree $%
n=M/2-S^{z}$. These solutions must correspond to regular Bethe states which
yield the highest weight vectors of the various $sl_{2}$ modules, while the
\textquotedblleft missing\textquotedblright\ states are simply descendant
states from highest weight vectors which lie in a different (higher)
spin-sector. For instance, in the case of even $M$ the possible number of
highest weight states in the sector $S^{z}=0$ is given by $\binom{M}{M/2}-%
\binom{M}{M/2-1}$ and we find that this number is matched by the solutions
to (\ref{PSwronski}); see the table below.{\small \medskip }

\begin{center}
\begin{tabular}{||c||c||c||c||c||c||c||c||c||}
\hline
$M$ & 3 & 4 & 5 & 6 & 7 & 8 & 9 & 10 \\ \hline
$S^{z}$ & 1/2 & 0 & 1/2 & 0 & 1/2 & 0 & 1/2 & 0 \\ \hline
dim & 3 & 6 & 10 & 20 & 35 & 70 & 126 & 252 \\ \hline
No & 2 & 2 & 5 & 5 & 14 & 14 & 42 & 42 \\ \hline
\end{tabular}%
{\small \medskip }

{\small Table 3. Number of solutions to (\ref{PSwronski}) in comparison with
the dimension of the spin sector.\medskip }
\end{center}

The simplified expression for the transfer matrix in terms of the two
linearly independent solutions $\mathcal{Q}^{\pm }$ remains formally the
same \cite{PS99}, however, we remind the reader that the degree of $\mathcal{%
Q}^{-}$ has changed in comparison with (\ref{twn}),%
\begin{equation}
\tilde{t}(u)=\mathcal{Q}^{-}(u+i)\mathcal{Q}^{+}(u-i)-\mathcal{Q}^{-}(u-i)%
\mathcal{Q}^{+}(u+i)\ .
\end{equation}%
Naturally, one wonders how the solutions $\mathcal{Q}^{\pm }$ are related to
the ones at quasi-periodic boundary conditions, $\tilde{Q}_{\omega }^{\pm
}(u)$, when the limit $\omega \rightarrow 1$ is taken. One finds that only a
subset of the solutions $Q_{\omega }^{\pm }$ stays finite, the other
solutions diverge. In the explicit construction of the $Q$-operator this is
due to the fact that the trace over the infinite-dimensional auxiliary space
does not converge any longer. The number of finite solutions, i.e. those for
which the limit $\omega \rightarrow 1$ is well-defined, approach the
solution $\mathcal{Q}^{+}$ of Pronko and Stroganov:%
\begin{equation}
\text{if\quad }\lim_{\omega \rightarrow 1}|\tilde{Q}_{\omega }^{\pm
}(u)|<\infty \text{\quad then\quad }\lim_{\omega \rightarrow 1}\tilde{Q}%
_{\omega }^{\pm }(u)=\mathcal{Q}^{+}(u)\ .
\end{equation}%
The above relation has been numerically verified for spin-chains up to
length $M=10$. Note that both solutions $Q_{\omega }^{+}$ and $Q_{\omega
}^{-}$ approach in the limit $\omega \rightarrow 1$ the same solution $%
\mathcal{Q}^{+}$ above the equator. This is to be expected as the degree $%
M-n $ of $Q_{\omega }^{-}$ can become smaller in the limit of periodic
boundary conditions but not greater. At the moment there appears to be no $Q$%
-operator construction which would yield the other solution $\mathcal{Q}^{-}$
and at the same time have the analogous factorization property (\ref{Qfactor}%
). The constructions suggested in the literature for periodic boundary
conditions \cite{De99,Pr00,De05} all have degree $\leq M$ for the spin 1/2
chain of $M$ sites, while the maximal degree of $\mathcal{Q}^{-}$ is $M+1$%
.\medskip

\begin{center}
\begin{tabular}{|c|c|c|}
\hline\hline
$i\mathcal{Q}^{-}(u)$ & $\mathcal{Q}^{+}(u)$ & $\tilde{t}(u)$ \\ \hline\hline
$u^{4}-\frac{3}{2}u^{2}-\frac{1}{48}$ & $u(u+\frac{1}{4})$ & $-\frac{25}{32}+%
\frac{15}{8}u^{2}+\frac{9}{2}u^{4}+2u^{6}$ \\ \hline\hline
$u^{4}+\frac{4\mp \sqrt{13}}{2}u^{2}-\frac{7\mp 2\sqrt{13}}{16}$ & $u^{3}+%
\frac{5\mp 2\sqrt{13}}{12}u$ & $\frac{31\pm 8\sqrt{13}}{32}+\frac{7\pm 8%
\sqrt{13}}{8}u^{2}+\frac{9}{2}u^{4}+2u^{6}$ \\ \hline\hline
$u^{4}+u^{2}\pm \frac{u}{2\sqrt{3}}-\frac{1}{16}$ & $u^{3}+\frac{u}{12}\pm 
\frac{1}{4\sqrt{3}}$ & $-\frac{1}{32}\mp \sqrt{3}u+\frac{23}{8}u^{2}+\frac{9%
}{2}u^{4}+2u^{6}$ \\ \hline\hline
\end{tabular}%
\medskip
\end{center}

\noindent {\small Table 4. Solutions to the quantum Wronskian (\ref%
{PSwronski}) for }${\small M=6,~S}^{z}{\small =0}${\small \ and the
corresponding eigenvalues of the transfer matrix.}

\section{Conclusions}

In this work we have presented the isotropic limit of a previous Q-operator
construction for the XXZ model \cite{CKQ2,CKQ3,CKQ4,CKQ7} in order to
discuss the XXX model with quasi-periodic boundary conditions. The
motivation for this discussion has been twofold. On the one hand this
construction enables one to formulate an analytic continuation of the fusion
hierarchy to complex dimension as it has been recently used in the
description of correlation functions in form of a trace functional \cite%
{BJMSTxxx}. In this context it should be noted that previous constructions
of Q-operators for the XXX model \cite{De99,Pr00,De05} have always been for
periodic boundary conditions where an analogous formulation does not exist.
This is due to the fact that the trace over an infinite-dimensional
auxiliary space has to be taken whose convergence is not necessarily
guaranteed. Moreover, due to the $sl_{2}$ symmetry the set of solutions to
the Bethe ansatz equations is reduced (i.e. only the highest weight states
in each $sl_{2}$ module are proper Bethe states), whence certain functional
relations such as the quantum Wronskian for periodic boundary conditions 
\cite{PS99} do not yield the complete set of eigenvalues; compare with table
3.

This provided additional motivation for investigating a Q-operator for the
twisted XXX model. Via this construction one is lead to a quantum Wronskian
for quasi-periodic boundary conditions (see (\ref{Wronski}) in the text),
which now yields the complete set of Bethe states and eigenvalues of the
transfer matrix. Our derivation relied on previous algebraic Bethe ansatz
results for the Q-operator of the XXZ model \cite{CKQ3}. As emphasized in
the text the quantum Wronskian has a simpler structure than the Bethe ansatz
equations and based on numerical computations we found special solutions for
spin-chains of even length and vanishing total spin satisfying more
fundamental identities. For instance the Bethe roots of the aforementioned
subset of solutions obey the set of equations,%
\begin{equation*}
(v_{j}^{+}+i/2)^{M}=\frac{\omega ^{-1}}{\omega ^{-1}-\omega }%
~\tprod_{k=1}^{M/2}(v_{j}^{+}-v_{k}^{+}+i)(v_{j}^{+}+v_{k}^{+})
\end{equation*}%
and are either real or occur in complex conjugate pairs; see the discussion
in section 5.2. Among these special solutions is the eigenvalue which
corresponds to the groundstate in the limit of periodic boundary conditions
and has real Bethe roots. The present numerical data only include chains up
to length $M=10$ and further investigation is needed to see whether they
persist for longer chains. This is particular important in order to make
contact with the thermodynamic Bethe ansatz and the string hypothesis \cite%
{Bethe} \cite{T71} \cite{FT81}. As it has been discussed in the literature
there might be a critical length beyond which certain solutions cease to
exist, see e.g. \cite{Bethe}\ \cite{EVS92}. We leave this problem of a more
extensive numerical study to future work.\bigskip 

\noindent \textbf{Acknowledgments: }This article has been motivated by
discussions at the DFG Summer School "Representation Theory in Mathematical
Physics", 18-22 July 2005, Bad Honnef, Germany and the author would like to
thank the organizers and the participants for interesting discussions. This
work is financially supported by a University Research Fellowship of the
Royal Society.


\begin{thebibliography}{10}

\bibitem{Bx72}
R.~J. Baxter,
\newblock Partition Function of the Eight-Vertex Lattice Model,
\newblock Ann. Phys., NY {\bf 70}, 193--228 (1972).

\bibitem{Bx73}
R.~J. Baxter,
\newblock Eight-Vertex Model in Lattice Statistics and One-Dimenisional
  Anisotropic Heisenberg Chain I-III,
\newblock Ann. Phys., NY {\bf 76}, 1--24,25--47,48--71 (1973).

\bibitem{Bx82}
R.~J. Baxter,
\newblock Exactly Solved Models in Statistical Mechanics,
\newblock Academic Press, London  (1982).

\bibitem{PG}
V.~Pasquier and M.~Gaudin,
\newblock The periodic Toda chain and a matrix generalization of the Bessel
  function recursion relations,
\newblock J. Phys. A: Math. Gen. {\bf 25}, 5243--5252 (1992).

\bibitem{KLWZ97}
I.~Krichever, O.~Lipan, P.~Wiegmann, and A.~Zabrodin,
\newblock Quantum integrable models and discrete classical Hirota equations,
\newblock Comm. Math. Phys. {\bf 188}, 267--304 (1997).

\bibitem{BLZ97}
V.~Bazhanov, S.~Lukyanov, and A.~Zamolodchikov,
\newblock Integrable structure of conformal field theory II: Q-operator and DDV
  equation,
\newblock Comm. Math. Phys. {\bf 190}, 247--278 (1997).

\bibitem{KS98}
V.~Kuznetsov and E.~K. Sklyanin,
\newblock On B\"{a}cklund transformations for many-body systems,
\newblock J. Phys. A: Math. Gen. {\bf 31}, 22241--2251 (1998).

\bibitem{Sk00}
E.~K. Sklyanin,
\newblock B\"{a}cklund transformations and Baxter's Q-operator, in 'Integrable
  systems: from classical to quantum', Montreal, QC, 1999,
\newblock CRM Proc. Lecture Notes {\bf 26}, 227--250 (AMS, Providence, RI,
  2000).

\bibitem{FKV01}
L.~D. Faddeev, R.~Kashaev, and A.~Y. Volkov,
\newblock Strongly Coupled Quantum Discrete Liouville Theory I: Algebraic
  Approach and Duality,
\newblock Comm. Math. Phys. {\bf 219}, 199--219 (2001).

\bibitem{Bethe}
H.~Bethe,
\newblock Zur Theorie der Metalle I. Eigenwerte und Eigenfunktionen der
  linearen Atomkette,
\newblock Z. Physik {\bf 71}, 205--226 (1931).

\bibitem{Li67}
E.~H. Lieb,
\newblock Exact Solution of the Two-Dimensional Slater KDP Model of a
  Ferroelectric,
\newblock Phys. Rev. Lett. {\bf 19}, 108--110 (1967).

\bibitem{Su67}
B.~Sutherland,
\newblock Exact Solution of a Two-Dimensional Model for Hydrogen-Bonded
  Crystals,
\newblock Phys. Rev. Lett. {\bf 19}, 103--104 (1967).

\bibitem{FM01}
K.~Fabricius and B.~M. McCoy,
\newblock Bethe's equation is incomplete for the XXZ model at roots of unity,
\newblock J. Stat. Phys. {\bf 103}, 647--678 (2001).

\bibitem{Bx02}
R.~J. Baxter,
\newblock Completeness of the Bethe ansatz for the six and eight-vertex models,
\newblock J. Stat. Phys. {\bf 108}, 1--48 (2002).

\bibitem{DFM}
T.~Deguchi, K.~Fabricius, and B.~M. McCoy,
\newblock The $sl_2$ loop algebra symmetry of the six-vertex model at roots of
  unity,
\newblock J. Stat. Phys. {\bf 102}, 701--736 (2001).

\bibitem{FT81}
L.~A. Takhtajan and L.~D. Faddeev,
\newblock Spectrum and scattering of stimuli of excitations in the
  one-dimensional isotropic Heisenberg model,
\newblock Zap. Nauch. Semin LOMI {\bf 109}, 134 (1981).

\bibitem{Deg01}
T.~Deguchi,
\newblock Non-regular eigenstate of the XXX model as some limit of the Bethe
  state,
\newblock J. Phys. A: Math. Gen. {\bf 34}, 9755--9775 (2001).

\bibitem{BY61}
N.~Byers and C.~Yang,
\newblock Theoretical Considerations Concerning Quantized Magnetic Flux in
  Superconducting Cylinders,
\newblock Phys. Rev. Lett. {\bf 7}, 46--49 (1961).

\bibitem{dV84}
H.~J. de~Vega,
\newblock Families of commuting transfer matrices and integrable models with
  disorder,
\newblock Nucl. Phys. B {\bf 240}, 495--513 (1984).

\bibitem{ABB88}
F.~C. Alcaraz, M.~N. Barber, and M.~T. Batchelor,
\newblock Conformal Invariance, the XXZ Chain and the Operator Content of
  Two-Dimensional Critical Systems,
\newblock Ann. Phys., NY {\bf 182}, 280--343 (1988).

\bibitem{YF92}
N.~Yu and M.~Fowler,
\newblock Twisted boundary conditions and the adiabatic ground state for the
  attractive XXZ Luttinger liquid,
\newblock Phys. Rev. B {\bf 46}, 14583--14593 (1992).

\bibitem{FST79}
L.~D. Faddeev, E.~K. Sklyanin, and L.~A. Takhtajan,
\newblock Quantum Inverse problem I,
\newblock Theor. Math. Phys. {\bf 40}, 194--220 (1979).

\bibitem{TV95}
V.~Tarasov and A.~Varchenko,
\newblock Bases of Bethe Vectors and Difference Equations with Regular Singular
  Points,
\newblock Int. Math. Res. Notices {\bf 13}, 637--669 (1995).

\bibitem{PS99}
G.~P. Pronko and Y.~G. Stroganov,
\newblock Bethe equations on the wrong side of equator,
\newblock J. Phys. A: Math. Gen. {\bf 32}, 2333--2340 (1999).

\bibitem{De99}
S.~E. Derkachov,
\newblock Baxter's Q-operator for the homogeneous XXX spin chain,
\newblock J. Phys. A: Math. Gen. {\bf 32}, 5299--5316 (1999).

\bibitem{Pr00}
G.~P. Pronko,
\newblock On the Baxter's Q-operator for the XXX spin chain,
\newblock Comm. Math. Phys. {\bf 212}, 687--701 (2000).

\bibitem{De05}
S.~E. Derkachov,
\newblock Factorization of the R-matrix and Baxter's Q-operator,
\newblock math.QA/0507252 .

\bibitem{DKK05}
S.~E. Derkachov, D.~Karakhanyan, and R.~Kirschner,
\newblock Baxter's Q-operators of the XXZ chain and R-matrix factorization,
\newblock hep-th/0511024 .

\bibitem{BJMSTxxx}
H.~Boos, M.~Jimbo, T.~Miwa, F.~Smirnov, and Y.~Takeyama,
\newblock A recursion formula for the correlation functions of an inhomogeneous
  XXX model,
\newblock hep-th/0405044 .

\bibitem{BJMSTxxz}
H.~Boos, M.~Jimbo, T.~Miwa, F.~Smirnov, and Y.~Takeyama,
\newblock Reduced qKZ equation and correlation functions of the XXZ model,
\newblock hep-th/0412191 .

\bibitem{BJMSTxyz}
H.~Boos, M.~Jimbo, T.~Miwa, F.~Smirnov, and Y.~Takeyama,
\newblock Traces on the Sklyanin algebra and correlation functions of the
  eight-vertex model,
\newblock J. Phys. A: Math. Gen. {\bf 38}, 7629--7660 (2005).

\bibitem{BJMSTxxx2}
H.~Boos, M.~Jimbo, T.~Miwa, F.~Smirnov, and Y.~Takeyama,
\newblock Density matrix of a finite sub-chain of the Heisenberg
  anti-ferromagnet,
\newblock hep-th/0506171 .

\bibitem{CKQ3}
C.~Korff,
\newblock Auxiliary matrices for the six-vertex model and the algebraic Bethe
  Ansatz,
\newblock J. Phys. A: Math. Gen. {\bf 37}, 7227--7253 (2004).

\bibitem{CKQ7}
C.~Korff,
\newblock A Q-operator identity for the correlation functions of the infinite
  XXZ spin-chain,
\newblock J. Phys. A: Math. Gen. {\bf 38}, 6641--6657 (2005).

\bibitem{RW02}
M.~Rossi and R.~Weston,
\newblock A generalized Q-operator for $U_q(\hat{sl}_2)$ vertex models,
\newblock J. Phys. A: Math. Gen. {\bf 35}, 10015--10032 (2002).

\bibitem{CKQ2}
C.~Korff,
\newblock Auxiliary matrices for the six-vertex model at roots of unity II.
  Bethe roots, complete strings, and the Drinfeld polynomial,
\newblock J. Phys. A: Math. Gen. {\bf 37}, 385--406 (2004).

\bibitem{CKQ4}
C.~Korff,
\newblock Auxiliary matrices on both sides of the equator,
\newblock J. Phys. A: Math. Gen. {\bf 38}, 47--67 (2005).

\bibitem{T71}
M.~Takahashi,
\newblock One-Dimensional Heisenberg Model at Finite Temperature,
\newblock Prog. Theor. Phys. {\bf 46}, 401--415 (1971).

\bibitem{EVS92}
F.~H.~L. Essler, V.~E. Korepin, and K.~Schoutens,
\newblock Fine structure of the Bethe ansatz for the spin-1/2 Heisenberg XXX
  model,
\newblock J. Phys. A: Math. Gen. {\bf 25}, 4115--4126 (1992).

\end{thebibliography}
\end{document}